\documentclass[twocolumn]{aastex631}

\begin{document}

\title{GradFlex: a Gradient-Based Method for Weak Lensing Flexion Measurement}

\author[0000-0001-7169-4642]{Jiarui Sun}
\affiliation{Tsung-Dao Lee Institute, Shanghai Jiao Tong University, Shanghai 201210, China}
 \affiliation{State Key Laboratory of Dark Matter Physics, School of Physics and Astronomy, \\
Shanghai Jiao Tong University, Shanghai 200240, China}%Lines break automatically or can be forced with \\
\author[0000-0003-0002-630X]{Jun Zhang}%
 \email{Contact author: betajzhang@sjtu.edu.cn}
 \affiliation{State Key Laboratory of Dark Matter Physics, School of Physics and Astronomy, \\
Shanghai Jiao Tong University, Shanghai 200240, China}
\author{Toshifumi Futamase}
\affiliation{Astronomical Institute, Tohoku University, 6-3 Aramaki, Aoba-ku, Sendai 980-8578, Japan}



\begin{abstract}

Flexion, the second-order gravitational lensing distortion, probes small-scale mass distribution and complement the cosmic shear measurement. However, its detection is highly challenging due to the faint signal and the contamination arising from the intrinsic galaxy morphology. In this Letter, we present GradFlex,  a new method that extracts flexion from the gradient field of galaxy surface brightness distributions without assuming a specific point-spread function model or galaxy morphology. Using surface-brightness conservation, we relate the pre- and post-lensing gradient fields and derive flexion estimators from ensemble-averaged statistics of the corresponding spins. Tests on numerical simulations across diverse galaxy morphologies and multiple noise realizations demonstrate its reliability under realistic observing conditions. Combined with shear measurements, it could tighten constraints on cluster density profiles and enable more detailed studies of substructure than shear alone. 

\end{abstract}



\section{Introduction}

Weak gravitational lensing shear measurement is well-established and widely used in precision cosmology and galaxy cluster mass reconstruction \citep{1995ApJ...449..460K,1997ApJ...475...20L,2011JCAP...11..041Z,2003ApJ...598..804K,2014ApJ...795..163U,2022PhRvD.105b3515S,2023PhRvD.108l3518L,2024SCPMA..6770413L,2024MNRAS.529..802H,2025arXiv250319441W,2025arXiv250707629S}. Conventional shear measurement techniques are generally optimized for the weak-distortion regime. However, growing attention has recently been paid to the challenges these methods face in regions where lensing signals become significant, motivating the development of methods that remain accurate under large distortions \citep{2025ApJ...992L..22S,2026MNRAS.546f2279O}.

Beyond shear, gravitational lensing produces the next order distortions known as flexion. Flexion provides complementary information about the lens mass distribution through its sensitivity to small-scale density variations \citep{2006MNRAS.365..414B}. Using simulated galaxy clusters, \citet{2016MNRAS.463.4287C} showed that adding flexion to joint strong-lensing and weak-lensing shear analyses enables the detection of lower-mass subhalos at finer angular scales, improving the spatial resolution and accuracy of cluster mass reconstructions. On galaxy scales, \citet{2011MNRAS.412.2665V} combined flexion and shear measurements from the Hubble Space Telescope Cosmic Evolution Survey (COSMOS) to constrain the average density profile of galaxy dark matter halos, obtaining tighter constraints than with shear alone.

However, measuring flexion reliably from galaxy images remains challenging, and existing methods rely on different assumptions and face distinct limitations. 
Shapelet-based methods expand galaxy images in an orthogonal basis. The flexion estimators described by \citet{2007ApJ...660.1003G} are confined to the weak field that retains only terms linear in the lensing fields. In practice, recovering higher-order distortions requires a choice of basis truncation
that balances morphological information against noise. A finite expansion can also introduce model mismatch when the galaxy profile is poorly represented by the chosen basis \citep{2010AA...510A..75M}. Moment-based approaches, such as HOLICs \citep{2008ApJ...680....1O}, also suffer from severe noise contamination in higher-order moments, resulting in highly noisy spin-3 HOLICs measurements that limit their applicability to real data. 
The ERA method \citep{2022MNRAS.516..668O}, based on reconstructing a circular, zero-ellipticity image from the observed distorted image, significantly improves the measurement precision compared with HOLICs. Nevertheless, real-space flexion estimators remain sensitive to galaxy centroid determination. 
Model-fitting methods \citep{2021MNRAS.501.4103F}, on the other hand, rely on assumptions about the galaxy light profile and can be limited for galaxies with complex morphologies. These limitations highlight the major challenges in flexion measurement, including strong sensitivity to noise, PSF effects, and realistic galaxy morphology, together with the complexity of accounting for shear–flexion couplings and higher-order lensing terms. Motivated by the advantages of Fourier-space methods such as Fourier\_Quad \citep{2015JCAP...01..024Z} in handling PSF effects and noise bias in shear measurements, we explore a new approach based on Fourier transforms for flexion measurement.

In this paper, we present a new method for flexion measurement based on the gradients of galaxy surface brightness distributions.  \S\ref{sec:method} introduces the lensing formalism and develops the GradFlex estimator. \S\ref{sec:numtest} validates the method with simulations. \S\ref{sec:dis} discusses limitations and future directions.

\section{GradFlex: gradient-based flexion measurement}
\label{sec:method}

We first introduce the lens mapping and the properties of flexion following \citet{2006MNRAS.365..414B, 2008AA...485..363S}, then construct estimators from the surface-brightness gradient field.

\subsection{Lensing formalism and flexion}

Throughout this formulation, a two-component quantity $(x_1,x_2)$ is written in complex form as $x=x_1+ix_2$. The locally linearized lens equation is
\begin{equation}
    \beta=(1-\kappa)\theta-\gamma\theta^* \, .
\end{equation}
Here $\theta$ and $\beta$ denote local image-plane and source-plane coordinates, respectively. The convergence $\kappa=(\psi_{,11}+\psi_{,22})/2$ changes the apparent size and brightness of a background galaxy, while the shear $\gamma=(\psi_{,11}-\psi_{,22})/2+i\psi_{,12}$ describes anisotropic stretching. The projected lensing potential is denoted by $\psi$, with commas indicating derivatives with respect to the image coordinates.

As the lensing signal increases, higher-order effects become non-negligible. Even for an individual galaxy image, treating the convergence and shear as constants is no longer valid. It requires expanding the lensing equation to the second order: 
\begin{equation}
    \beta=(1-\kappa) \theta-\gamma \theta^*-\frac{1}{4} \mathcal{F}^* \theta^2-\frac{1}{2} \mathcal{F} \theta \theta^*-\frac{1}{4} \mathcal{G}\left(\theta^*\right)^2 \, . 
    \label{eq:lensing-eq-flexion}
\end{equation}
Defining the complex derivative $\partial\equiv\partial_1+i\partial_2$, the two flexion fields are
\begin{eqnarray}
\mathcal{F}\equiv\partial\kappa
 &=&\frac{1}{2}\left[(\psi_{,111}+\psi_{,122})
 +i(\psi_{,112}+\psi_{,222})\right] \, , \\
\mathcal{G}\equiv\partial\gamma
 &=&\frac{1}{2}\left[(\psi_{,111}-3\psi_{,122})
 +i(3\psi_{,112}-\psi_{,222})\right] \, .
\end{eqnarray}
The first flexion, $\mathcal{F}$, traces the spatial gradient of convergence and produces a skewness in the galaxy light distribution, accompanied by a displacement of its light centroid relative to the mapped source centroid. The second flexion, $\mathcal{G}$, describes a threefold distortion of the image. These distinct angular patterns correspond to spin-1 and spin-3, respectively. Their spin symmetries provide the basis for separating the two flexion signals in the gradient statistics below.

\subsection{Flexion Estimation}

\begin{figure}[t]
 \centering
 \includegraphics[width=\columnwidth,pagebox=cropbox]{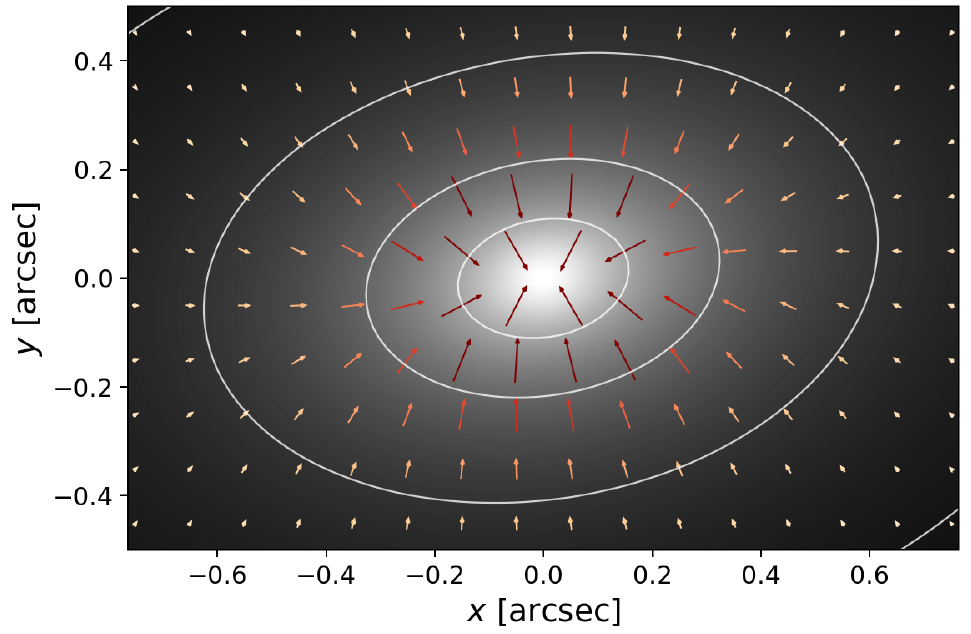}
 \caption{Surface-brightness gradient field of a simulated lensed galaxy. The surface-brightness distribution is indicated with the grayscale, and the corresponding gradient is shown as the arrows. }
 \label{fig:gradient-field}
\end{figure}

The surface-brightness gradient field also carries information about the shape of a galaxy, as illustrated in Figure~\ref{fig:gradient-field}. The directions and magnitudes of the gradient vectors describe how the brightness varies across the image, tracing its morphology and lensing signal. The following text shows how to compute the gradient field and relate its statistics to the flexion signals.

PSF effects and brightness gradients can be handled within the same Fourier-space framework. Dividing by the Fourier transform of the instrumental PSF and multiplying by that of an isotropic Gaussian replace the instrumental PSF with a Gaussian. Spatial derivatives then follow from multiplication by the corresponding wavevector components. We denote the resulting Gaussian-smoothed lensed image by $f_O$.
Using surface-brightness conservation and the local lens mapping, we derive the first-order relation between the gradient of $f_O$ and that of the intrinsic brightness convolved with the same Gaussian kernel. This relation separates the contributions from convergence, shear, and flexion, including the effects of Gaussian smoothing. The detailed derivation is presented in Appendix~\ref{sec:appendix}.
The complex brightness gradient has spin-1. Squaring the complex brightness gradient and multiplying it by its complex conjugate produces a spin-1 statistic, while cubing it produces a spin-3 statistic, matching the first and second flexions, respectively. Spatial integration followed by averaging over statistically isotropic intrinsic orientations removes terms with unmatched spin, leaving signals proportional to the corresponding flexion signals. Their amplitudes also depend on the galaxy brightness distribution and Gaussian smoothing. The spin-0 response factors $R_F$ and $R_G$ account for this dependence and provide the normalization for the estimators:
\begin{eqnarray}
 {\mathcal{F}}&=&\left\langle
 (\partial_1f_O+i\partial_2f_O)^2
 (\partial_1f_O-i\partial_2f_O)
 \right\rangle/R_F \\ \nonumber
 {\mathcal{G}}&=&\left\langle
 (\partial_1f_O+i\partial_2f_O)^3
 \right\rangle/R_G .
\end{eqnarray}
Here the brackets denote spatial integration followed by ensemble averaging. The response factors involve derivatives of the Gaussian-smoothed intrinsic brightness, image coordinates, and the smoothing-scale. The full response factors are given in Appendix~\ref{sec:appendix}.

\section{The Numerical Test}
\label{sec:numtest}
We test GradFlex using simulated galaxies with both regular and irregular morphologies. Following the point-source construction of \citet{2025ApJ...992L..22S}, we generate regular galaxies by placing points uniformly within a circular region and assigning their fluxes according to a S\'ersic profile,
\begin{equation}
 I(R)=I_e\exp\left\{-b_n\left[\left(\frac{R}{R_e}\right)^{1/n}-1\right]\right\} \, ,
\end{equation}
where $R_e$ is the half-light radius, $n$ is the S\'ersic index,  $I_e$ is the surface brightness at $R_e$, and $b_n = 1.999n - 0.327$ for $n \ge 1$ \citep{1989woga.conf..208C}. A linear transformation of the point coordinates introduces an intrinsic ellipticity of $e=0.2$. We vary $R_e$ and $n$ to test the dependence on the radial light profile. For irregular galaxies, the point positions are generated by a two-dimensional random walk with 1000 steps and a threshold distance of 8 pixels, producing asymmetric light distributions.

The input lensing fields are drawn from an NFW halo with $M_{200}=1.2\times10^{15}\,M_\odot$ and concentration $c=3$ \citep{1996ApJ...462..563N}, assuming lens redshift of $z_l=0.3$ and source redshift of $z_s=2$. We sample different angular separations from the halo center. At each separation, the halo supplies the corresponding convergence, shear, and flexion, with smaller separations probing stronger lensing distortions.  

After applying the lens mapping, we convolve the galaxies with a Moffat PSF of scale radius $\alpha=3$ pixels and power-law index $\beta=4$. Both noiseless and noisy simulations use image stamps of $128\times128$ pixels. The adopted pixel scale of $0.02^{\prime\prime}$ per pixel matches that of the drizzled NIRCam short-wavelength mosaics from the UNCOVER survey \citep{2024ApJS..270....7W,2024ApJ...974...92B}. In the noiseless tests, each galaxy is represented by 12 rotated copies separated by $30^\circ$ to suppress intrinsic shape contributions. Each noisy measurement combines $2.4\times10^6$ randomly oriented simulated galaxies.

Although our estimators are derived to first order in the lensing fields, simulations without shear show that they recover the reduced flexions, $F\equiv\mathcal{F}/(1-\kappa)$ and $G\equiv\mathcal{G}/(1-\kappa)$, analogous to the reduced shear. This behavior holds across the galaxy morphologies tested. We therefore interpret the measured signals as reduced flexion and compare them with the corresponding input values below.

\subsection{Measurement without Noise}
\label{sec:Measurement without Noise}
\begin{figure}
    \centering
    \includegraphics[width=0.5\textwidth]{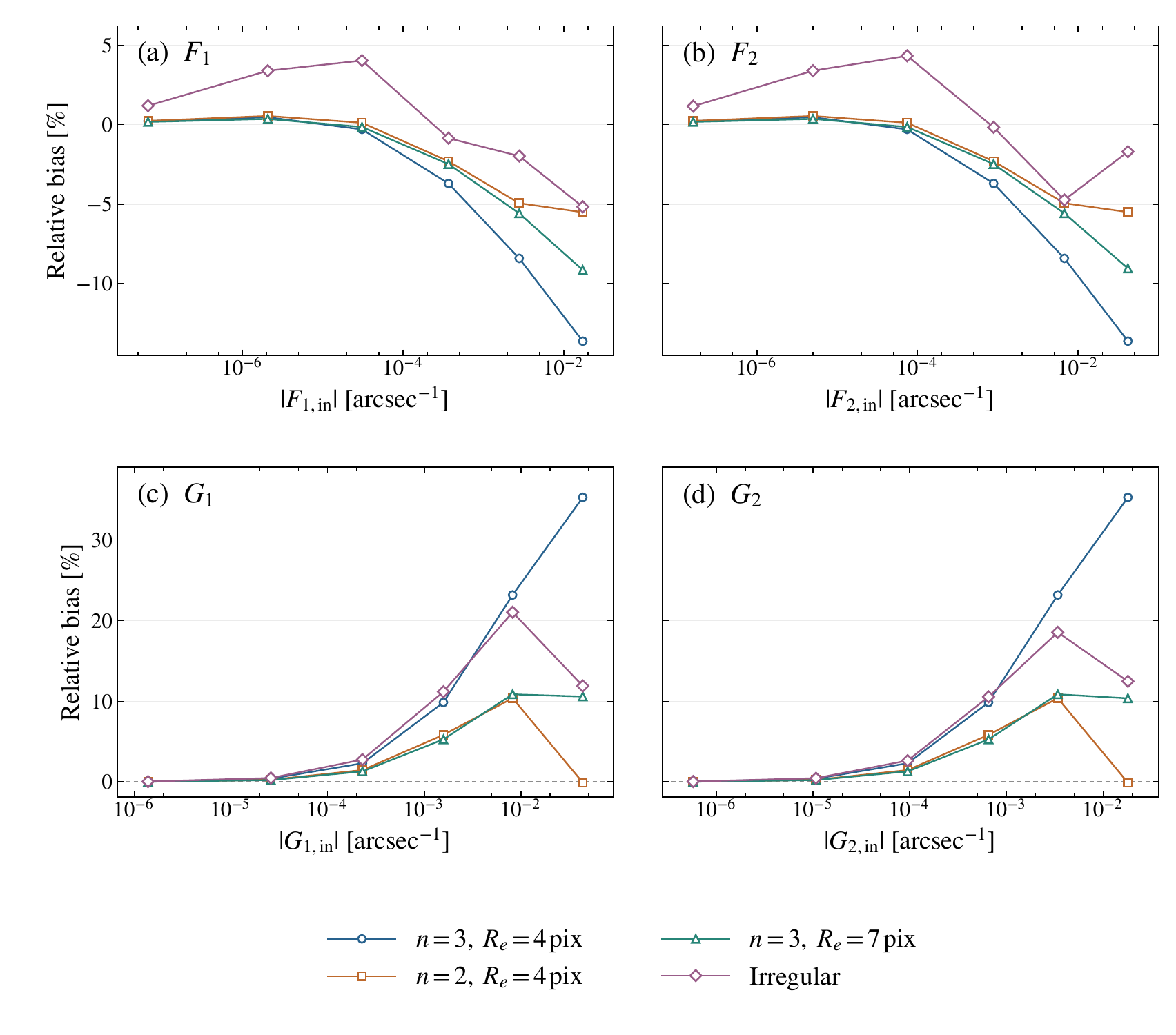}
    \caption{Relative biases in noiseless flexion measurements as a function of the input reduced-flexion components for the simulated galaxy morphologies. The first and second rows show results for the first and second reduced flexions, respectively. The input signals are evaluated at angular separations of $4000$, $1200$, $428$, $140$, $45$, and $15\,\mathrm{arcsec}$ from the NFW halo center, corresponding to the plotted points from the left to the right. The corresponding input reduced-shear amplitude $|g|$ ranges from approximately $10^{-3}$ to $0.35$. The blue, green and orange lines correspond to regular galaxies, with the corresponding effective radii and S\'ersic index as shown in the text at the bottom of the figure. The purple line is for irregular galaxies.}
    \label{fig:moffat-nonoise}
\end{figure}

Figure~\ref{fig:moffat-nonoise} shows the relative measurement biases as a function of the input reduced-flexion components for different galaxy morphologies. The first and second rows correspond to the first and second reduced flexions, respectively. At small input signals, the recovered flexions closely match the input values across the tested morphologies, demonstrating the accuracy of GradFlex in the weak-distortion regime.

As the lensing strength increases toward the halo center, the measurements show larger systematic deviations from the input values, reflecting the higher-order terms neglected in the first-order derivation. 
The bias also depends on the source morphology. Under the same lensing conditions, regular galaxies with different $n$ and $R_e$ exhibit different bias levels. This leads to a more complicated correction of the bias. Possible mitigation strategies are discussed in \S\ref{sec:dis}.

\subsection{Measurement with Noise}
We compare noisy measurements with the noiseless reference to assess additional noise-induced bias. Agreement with this reference does not imply unbiased recovery of the input flexion, because the systematic offsets discussed in Section~\ref{sec:Measurement without Noise} remain. 

\begin{figure}
    \centering
    \includegraphics[width=0.5\textwidth]{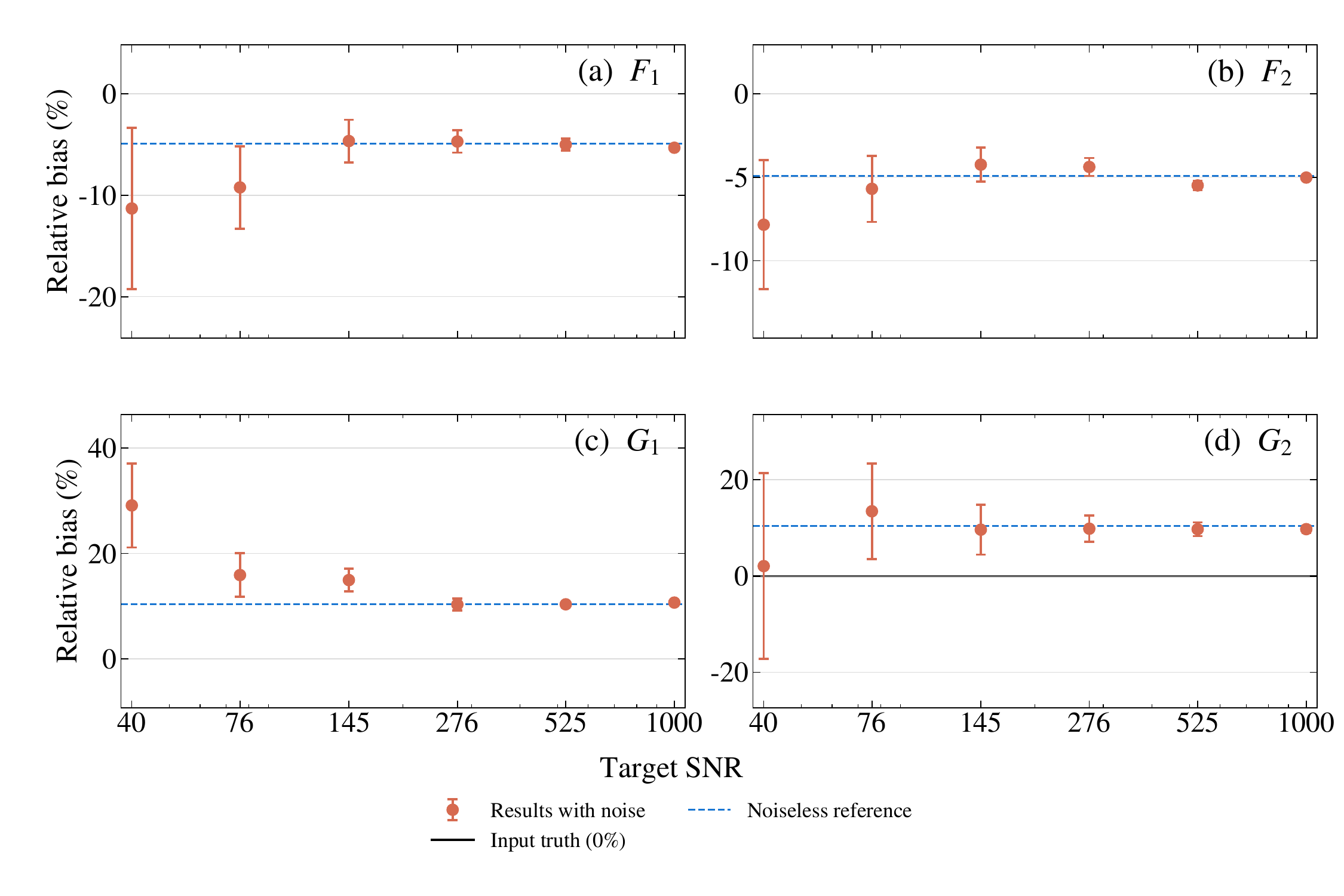}
    \caption{Relative biases as a function of SNR for regular galaxies with $R_e=4$ pixels, $n=2$, and $e=0.2$, at $45^{\prime\prime}$ from the NFW halo center. The input lensing fields are identical to those at the same $45^{\prime\prime}$ position in Fig.~\ref{fig:moffat-nonoise}. Red points show noisy measurements, each combining $2.4\times10^6$ randomly oriented galaxies. Blue dashed lines indicate the noiseless reference at the same input lensing fields, and black lines mark zero bias. The top and bottom rows show the first- and second-flexion components, respectively. }
    \label{fig:moffat-noise}
\end{figure}

Figure~\ref{fig:moffat-noise} shows the relative biases of the four reduced-flexion components as a function of SNR at a fixed halo-centric separation of $45^{\prime\prime}$, using the circular mask defined from the FOF-identified region as described below. All four components agree with the noiseless reference within $3\sigma$ over the range shown. Here $\sigma$ denotes the standard error estimated using a jackknife with $10^5$ groups. Tests at halo-centric separations of $15^{\prime\prime}$ and $75^{\prime\prime}$ yield consistent conclusions.

For each noisy image, we construct a circular mask based on a Friends-of-Friends (FOF) selection. The FOF algorithm identifies the region connected to the brightest pixel through pixels with intensities exceeding one standard deviation of the background noise, allowing connections along shared edges or corners. The circular mask is centered on the brightest pixel, with its radius set by the distance to the most distant pixel in the FOF-identified region. The mask defines the region for the final spatial integrals and for calculating the flux-weighted centroid, which sets the coordinate origin for the response factors. Note that the circular mask itself remains centered on the brightest pixel. In addition, brightness derivatives of all orders are computed from the full image using Fourier transforms before the mask is applied.

We compare measurements using this FOF-based circular mask with those restricted to the FOF-identified region itself. For the latter, first-flexion estimates increasingly deviate from the noiseless reference as SNR decreases, with discrepancies clearly exceeding $3\sigma$ at low SNR. No comparable trend is detected with the FOF-based circular mask, without additional treatment of the background noise. These results favor retaining faint galaxy light rather than restricting measurements to the FOF-identified region. At low SNR, including more peripheral background noise increases the statistical uncertainties, but no systematic shift is detected in these tests.  

We also test sensitivity to the choice of coordinate origin. For regular galaxies, using either the brightest pixel or the flux-weighted centroid as the origin changes the recovered signals by less than $2\%$, even at $\mathrm{SNR}\sim40$. At lower SNR, noise can affect both mask construction and galaxy centering. The brightest pixel may correspond to a background noise peak, leading to an incorrectly identified FOF region and coordinate origin. In our tests, these procedures remain stable for images with \(\mathrm{SNR}>20\). For irregular galaxies, differences can reach $5\%$ for individual objects, with first flexion becoming more sensitive as morphological asymmetry increases. Second flexion remains comparatively insensitive to centering. These tests indicate that either the brightest pixel or the flux-weighted centroid can be used as the coordinate origin, with only a small effect on the recovered signals. This conclusion assumes that the galaxy can be reliably identified and centered. 

\section{Summary and Discussion} \label{sec:dis}

As a second-order weak-lensing distortion, flexion is intrinsically faint and considerably more challenging to measure than shear. We have developed a perturbation-theory-based formalism that estimates flexion from gradients of galaxy surface brightness. Simulation tests incorporating PSF effects and realistic noise demonstrate that the method performs well under typical observational conditions.

GradFlex differs from conventional moment-based flexion estimators in the image statistics it uses. Moment-based approaches construct estimators from surface-brightness multipole moments, whereas GradFlex uses products of brightness gradients. In HOLICs, normalization involves fourth-order coordinate moments, while weighted HOLICs requires even higher-order coordinate moments to account for the weight function. Such coordinate factors can enhance the contribution of noise at large radii. GradFlex also contains moment-like quantities: its response factors include coordinate-weighted integrals of products of brightness derivatives. However, the explicit coordinate factors are at most quadratic, and the terms quadratic in position contribute only a small fraction of the total responses in our tests. This is consistent with the limited sensitivity to peripheral background noise and centering found.

This framework isolates distortions with specific spin symmetry and can be extended to construct estimators for higher-order lensing distortions of different spin.  Moreover, idealized regular galaxies have centrosymmetric intrinsic brightness distributions, free of intrinsic flexion-like asymmetry. In principle, this allows flexion to be recovered from individual regular galaxies rather than relying solely on ensemble averaging. The use of Fourier Transfer based operations to construct the gradient field also provides a convenient framework for incorporating the effects of the PSF, since convolution in real space can be treated straightforwardly in Fourier space. 

Despite these advantages, the method has several limitations. The current implementation relies on several perturbative approximations: mapping the lens equation from image to source plane, linearizing the Jacobian, and substituting the unlensed surface brightness with its lensed counterpart. These operations omit higher-order terms such as $\gamma^2$ and $\gamma\mathcal{F}$, which become significant in large-shear regimes (e.g., cluster centers). The resulting calibration issues will be addressed in our next work. It will also examine weighting schemes for samples spanning a range of SNRs and assess contamination of the brightness gradients by foreground objects and neighboring sources. 

\begin{acknowledgments}
We thank Shuren Zhou for useful
suggestions. This work is supported by the National Key Basic Research and Development Program of China (2023YFA1607800, 2023YFA1607802), the NSFC grant (12573004), and the science research grants from China Manned Space Project (No. CMS\-CSST\-2021\-A01). J.S. is supported by T.D. Lee scholarship. The computations in this paper were run on the $\pi$ 2.0 cluster supported by the Center of High Performance Computing at Shanghai Jiaotong University.
\end{acknowledgments}

\appendix

\section{Detailed Derivation of GradFlex}
\label{sec:appendix}
We begin with the local lens mapping, 
\begin{equation}
 \beta_i=\theta_i-\psi_{,ij}\theta_j-\frac{1}{2}\psi_{,ijk}\theta_j\theta_k \, ,
\end{equation}
Here $\theta$ and $\beta$ denote local image-plane and source-plane coordinates, respectively. For brevity, we adopt the Einstein notation. When an index variable appears twice in a single item, it means that the item sums all the values of the index. To first order in the lensing-potential derivatives, the inverse mapping is
\begin{equation}
 \theta_i=\beta_i-A_{ij}\beta_j-\frac{1}{2}B_{ijk}\beta_j\beta_k \, ,
\end{equation}
with
\begin{equation}
 A_{ij}=-\psi_{,ij},\qquad B_{ijk}=-\psi_{,ijk} \, .
\end{equation}
This approximation requires the second derivative of the lensing potential to be small, meaning both the shear and convergence must be treated as small quantities. 

Let $f_S$ and $f_L$ denote the intrinsic and lensed surface-brightness distributions. Surface-brightness conservation gives $f_L(\vec{\theta})=f_S[\vec{\beta}(\vec{\theta})]$, linking the brightness distributions through the mapping above. As in \S\ref{sec:method}, we handle the instrumental PSF in Fourier space and use $f_O$ for the image with an isotropic Gaussian PSF:
\begin{equation}
 f_O(\vec{\theta})=\int d^2\vec{\theta}\,
 W(\vec{\theta}-\vec{\theta}^{\prime})f_L(\vec{\theta}^{\prime}) \, ,
\end{equation}
where $W(\vec{\theta}) \propto \exp[-|\vec{\theta}|^2/(2\sigma^2)]$, $\sigma$ is the standard deviation. $\vec{\theta}$ is the position at which the smoothed image is evaluated, and $\vec{\theta}'$ is the integration variable in the same image-plane coordinate system. 

To first order, the expression for $f_{O}\left(\vec{\theta}\right)$ is mathematically equivalent to: 
\begin{eqnarray}
f_{O}\left(\vec{\theta}\right)  &=& \int d^2\vec{\theta}\,
 W(\vec{\theta}-\vec{\theta}^{\prime})f_S(\vec{\beta}(\vec{\theta}^{\prime})) \\
&=&\int d^{2} \vec{\beta} \left|\det\frac{\partial\vec{\theta}'}{\partial\vec{\beta}}\right|  W\left(\vec{\theta}-\vec{\beta}\right)\left[1+\partial_{u} \ln W \cdot\left(A_{u \alpha} \beta_{\alpha}+B_{u \alpha w} \beta_{\alpha} \beta_{w} / 2\right)\right] f_{S}(\vec{\beta}) \, . \nonumber
\end{eqnarray}
Here and below, $\partial_u\equiv\partial/\partial\theta_u$ acts on the image plane. 
The first-order Jacobian determinant and the logarithmic derivative of the Gaussian kernel are 
$$ \left|\det\frac{\partial\vec{\theta}}{\partial\vec{\beta}}\right| =1-A_{11}-A_{22}-B_{22 v} \beta_{v}-B_{11 v} \beta_{v} \, , \quad \partial_{u} \ln W=-\frac{\theta_u-\beta_u}{ \sigma^{2}} \, .$$
Therefore,
\begin{eqnarray}
f_{O}\left(\vec{\theta}\right) &=  &\int d^{2} \vec{\beta} \cdot W\left(\vec{\theta}-\vec{\beta}\right) f_{S}(\vec{\beta}) \label{eq:fO} \\ 
& &\cdot\left[1-A_{11}-A_{22}-B_{22 v} \beta_{v}-B_{11 v} \beta_{v} -\frac{1}{\sigma^{2}}\left(\theta_{u}-\beta_{u}\right)\left(A_{u \alpha} \beta_{\alpha}+B_{u \alpha w} \beta_{\alpha} \beta_{w} / 2\right)\right] \, . \nonumber
\end{eqnarray}
We define the Gaussian-smoothed intrinsic brightness as $f_I(\vec{\theta})=\int d^2\vec{\beta}W(\vec{\theta}-\vec{\beta})f_S(\vec{\beta})$ and denote its derivatives with respect to $\vec{\theta}$ by $f_{I,ij\ldots k}\equiv\partial_i\partial_j\cdots\partial_k f_I$. Differentiating under the integral gives
$$f_{I,i}=-\frac{1}{\sigma^{2}} \int d^{2} \vec{\beta} \cdot\left(\theta_{i}-\beta_{i}\right) \cdot W(\vec{\theta}-\vec{\beta}) f_{S}(\vec{\beta})\, .$$
Therefore, it can be concluded:
\begin{equation}
    \int d^{2} \vec{\beta} \cdot \beta_{i} \cdot W(\vec{\theta}-\vec{\beta}) f_{S}(\vec{\beta})=\sigma^{2} f_{I,i}+\theta_{i} f_{I} \, . \label{eq:betaW}
\end{equation}
And repeated differentiation gives the second and third moments: 
\begin{eqnarray}
\label{eq:betabetaW}
\int d^{2} \vec{\beta} \cdot \beta_{i} \beta_{j} \cdot W(\vec{\theta}-\vec{\beta}) f_{S}(\vec{\beta})=\sigma^{4} f_{I,i j}+\delta_{i j} \sigma^{2} f_{I}+\theta_{i} \theta_{j} f_{I}+\sigma^{2}\left(\theta_{i} f_{I,j}+\theta_{j} f_{I,i}\right)\, , 
\end{eqnarray}
\begin{eqnarray}
\int d^{2} \vec{\beta} \cdot \beta_{i} \beta_{j} \beta_{k} \cdot W(\vec{\theta}-\vec{\beta}) f_{S}(\vec{\beta}) 
&=\theta_{i} \theta_{j} \theta_{k} f_{I}  +\sigma^{2}\left(\delta_{k j} \theta_{i} f_{I}+\delta_{i k} \theta_{j} f_{I}+\delta_{i j} \theta_{k} f_{I}+\theta_{i} \theta_{j} f_{I,k}+\theta_{i} \theta_{k} f_{I,j}+\theta_{j} \theta_{k} f_{I,i}\right) \nonumber \\
&+\sigma^{4}\left(\delta_{i j} f_{I,k}+\delta_{i k} f_{I,j}+\delta_{j k} f_{I,i}+\theta_{k} f_{I,i j}+\theta_{j} f_{I,i k}+\theta_{i} f_{I,k j}\right)+\sigma^{6} f_{I,i j k} \, . \nonumber
\end{eqnarray}

Substituting Eq.\ref{eq:betaW}-\ref{eq:betabetaW} into Eq.\ref{eq:fO}, the $f_{O}$ can be expressed as:
\begin{eqnarray}
f_{O}(\vec{\theta})  &=&  f_{I}[1  \left.-A_{11}-A_{22}\right]-\left(B_{22 i}+B_{11 i}+\sigma^{-2} A_{u i} \theta_{u}\right)\left(\sigma^{2} f_{I,i}+\theta_{i} f_{I}\right) \\
& & +\left(\sigma^{-2} A_{i j}-\frac{1}{2} \sigma^{-2} \theta_{u} B_{u i j}\right)\left[\sigma^{4} f_{I,i j}+\delta_{i j} \sigma^{2} f_{I}+\theta_{i} \theta_{j} f_{I} +\sigma^{2}\left(\theta_{i} f_{I,j}+\theta_{j} f_{I,i}\right)\right] \nonumber \\
& & +\frac{1}{2} \sigma^{-2} B_{i j k}\left[\theta_{i} \theta_{j} \theta_{k} f_{I}\right. +\sigma^{2}\left(\delta_{k j} \theta_{i} f_{I}+\delta_{i k} \theta_{j} f_{I}+\delta_{i j} \theta_{k} f_{I}+\theta_{i} \theta_{j} f_{I,k}+\theta_{i} \theta_{k} f_{I,j}+\theta_{j} \theta_{k} f_{I,i}\right) \nonumber \\
& & \left.+\sigma^{4}\left(\delta_{i j} f_{I,k}+\delta_{i k} f_{I,j}+\delta_{j k} f_{I,i}+\theta_{k} f_{I,i j}+\theta_{j} f_{I,i k}+\theta_{i} f_{I,k j}\right)+\sigma^{6} f_{I,i j k}\right] \, . \nonumber
\label{eq:f-F}
\end{eqnarray}
Therefore, 
\begin{eqnarray}
\partial_u f_{O} & = & f_{I,u}+A_{i j}  \left(\sigma^{2} f_{I,i j u}+\delta_{i u} f_{I,j}+\theta_{i} f_{I,j u}\right) \label{eq:partialf} \\
& & +\frac{1}{2} B_{i j k}\left[2 \delta_{i u} \theta_{j} f_{I,k}+\theta_{i} \theta_{j} f_{I,k u}+\sigma^{2}\left(\delta_{j k} f_{I,i u}+2 \delta_{j u} f_{I,i k}+2 \theta_{j} f_{I,i k u}\right) +\sigma^{4} f_{I,i j k u}\right] \nonumber \\ & = & f_{I,u}+\Gamma_{u k} f_{I,k}+  \frac{1}{2} B_{i j j} \sigma^{2} f_{I,i u}+B_{i k u} \sigma^{2} f_{I,i k}+\sigma^{2} \Gamma_{i k} f_{I,i k u}+\frac{1}{2} B_{i j k} \sigma^{4} f_{I,i j k u}+H_{k} f_{I,k u} \, , \nonumber 
\end{eqnarray}
where $\Gamma_{u k} = A_{k u}+\theta_{j} B_{u k j}, H_{u}  =  \theta_{i} A_{i u}+\theta_{i} \theta_{j} B_{u i j} / 2$.

To connect this result to the spin-based construction, we combine the brightness derivatives into complex fields $D_{ns}$, where the first index $n$ specifies the derivative order and the second index $s$ specifies the spin. The complex coordinate, higher-order brightness derivative fields, and lensing signals are defined as follows:
\begin{eqnarray}
\vartheta & = & \theta_{1}+i \theta_{2} \, , \label{eq:vartheta} \\
D_{11} & = & f_{I,1}+i f_{I,2} \, , \nonumber \\
D_{44} & = & \left(f_{I,1111}-6 f_{I,1122}+f_{I,2222}\right)+i\left(4 f_{I,1112}-4 f_{I,1222}\right) \, , \nonumber \\
D_{42} & = & \left(f_{I,1111}-f_{I,2222}\right)+2 i\left(f_{I,1211}+f_{I,1222}\right) \, , \nonumber \\
D_{40} & = & f_{I,1111}+2 f_{I,1122}+f_{I,2222} \, , \nonumber \\
D_{33} & = & \left(f_{I,111}-3 f_{I,221}\right)+i\left(3 f_{I,112}-f_{I,222}\right) \, , \nonumber \\
D_{22} & = & \left(f_{I,11}-f_{I,22}\right)+2 i f_{I,12} \, , \nonumber \\
D_{20} & = & f_{I,11}+f_{I,22} \, , \nonumber \\
D_{31} & = & \left(f_{I,111}+f_{I,221}\right)+i\left(f_{I,112}+f_{I,222}\right) \, , \nonumber \\
{\kappa} & = & - (A_{11}+A_{22}) /2 \, , \nonumber \\
{\gamma} & = & - (A_{11}-A_{22}+2 i A_{12})/2 \, , \nonumber \\
{\mathcal{F}} & = & - \left[ \left(B_{111}+B_{221}\right)+i\left(B_{112}+B_{222}\right) \right]/2 \, , \nonumber \\
{\mathcal{G}} & = & - \left[ \left(B_{111}-3 B_{221}\right)+i\left(3 B_{112}-B_{222}\right) \right] /2 \, . \nonumber 
\end{eqnarray}

With these definitions, Eq.~\ref{eq:partialf} becomes a complex gradient relation whose terms separate the effects of convergence, shear, and the two flexions:
\begin{eqnarray}
\partial_{1} f_{O}+i \partial_{2} f_{O} & = &
D_{11}-{\kappa} (D_{11}+\sigma^{2} D_{31} + \frac{1}{2} \vartheta D_{20} + \frac{1}{2} \vartheta^* D_{22}) \label{eq:ftoF} \\ \nonumber
& &-{\gamma} (D_{11}^*+\frac{1}{2} \sigma^{2} D_{31}^* +\frac{1}{2} \vartheta^* D_{20}) - \frac{1}{2} \hat{\gamma}^*(\sigma^{2} D_{33} + \vartheta D_{22}) \\ \nonumber 
& &- \frac{1}{2} {\mathcal{F}} (\vartheta D_{11}^*+\vartheta^* D_{11}+ \frac{1}{2} \sigma^{2} \vartheta D_{31}^* + \sigma^{2} \vartheta^* D_{31} + 3 \sigma^{2} D_{20} + \frac{3}{4} \sigma^{4} D_{40} + \frac{1}{2} \vartheta \vartheta^* D_{20} + \frac{1}{4} (\vartheta^*)^2 D_{22}) \\ \nonumber 
& &- {\mathcal{F}}^* (\sigma^{2} D_{22} + \frac{1}{2} \vartheta  D_{11} + \frac{1}{2} \sigma^{2} \vartheta  D_{31} + \frac{1}{4} \sigma^{2} \vartheta^*  D_{33} +  \frac{3}{8} \sigma^{4} D_{42} + \frac{1}{4} \vartheta \vartheta^* D_{22} + \frac{1}{8} \vartheta^2 D_{20}) \\ \nonumber 
& & - \frac{1}{2} {\mathcal{G}} (\sigma^{2} D_{22}^* + \vartheta^*  D_{11}^* + \frac{1}{2} \sigma^{2} \vartheta^*  D_{31}^* +  \frac{1}{4} \sigma^{4} D_{42}^*+ \frac{1}{4} (\vartheta^*)^2 D_{20})  \\ \nonumber
& & - \frac{1}{4} {\mathcal{G}}^* (\sigma^{2} \vartheta  D_{33} +  \frac{1}{2} \sigma^{4} D_{44}^*+ \frac{1}{2} \vartheta^2 D_{22}) \, . 
\end{eqnarray}

We construct a spin-1 statistic for the first flexion by squaring the gradient and multiplying it by its complex conjugate, and a spin-3 statistic for the second flexion by cubing the gradient. Substituting Eq.~\ref{eq:ftoF} into these statistics and retaining only terms linear in the lensing fields. Spatial integration followed by averaging over statistically isotropic intrinsic orientations eliminates nonzero-spin intrinsic coefficients, leaving terms proportional to the corresponding flexion signals. Their spin-0 coefficients define the response factors $R_F$ and $R_G$, giving  
\begin{eqnarray}
 \left\langle
 (\partial_1f_O+i\partial_2f_O)^2
 (\partial_1f_O-i\partial_2f_O)
 \right\rangle &=& {\mathcal{F}} R_F \\ \nonumber
 \left\langle
 (\partial_1f_O+i\partial_2f_O)^3
 \right\rangle &=&  {\mathcal{G}}R_G \label{eq:flexion-estimator}
\end{eqnarray}
Here the brackets denote spatial integration followed by ensemble averaging, and the response factors are 
\begin{eqnarray}
R_F&= &-\left\langle \left(\frac{5}{2}\vartheta^{*}D_{11}+\frac{3}{2}\sigma^{2} \vartheta^{*}D_{31}+3 \sigma^{2} D_{20} +\frac{3}{4} \sigma^{4} D_{40}+\frac{1}{2} \vartheta^{*} \vartheta D_{20}+\frac{1}{4} \vartheta^{* 2} D_{22}\right)|D_{11}|^{2}  \right.  \\ \nonumber  
& & +\left.\left(\frac{1}{2}\sigma^{2}\vartheta^{*}D_{31}^{*} +\frac{1}{4} \sigma^{2}\vartheta D_{33}^{*} +\sigma^{2} D_{22}^{*} +\frac{3}{8} \sigma^{4} D_{42}^{*} +\frac{1}{8} \vartheta^{* 2} D_{20}+\frac{1}{4} \vartheta^{*} \vartheta D_{22}^{*} \right)D_{11}^{2}\right\rangle \\ \nonumber 
R_G & = & -\frac{3}{2}\left\langle\left(\sigma^{2}D_{22}^*+\vartheta^*D_{11}^*+\frac{1}{2}\sigma^{2}\vartheta^*D_{31}^*+\frac{1}{4}\sigma^4D_{42}^*+\frac{1}{4}\vartheta^{*2}D_{20}  \right)D_{11}^2\right\rangle .
\label{eq:estimator-full}
\end{eqnarray}
The expressions for $R_F$ and $R_G$ involve derivatives of the Gaussian-smoothed intrinsic brightness, image coordinates, and the smoothing scale. To evaluate these factors from observations, we replace the unknown $f_I$ and its derivatives by $f_O$ and its derivatives. Equation~\ref{eq:f-F} shows that their difference is first order in the lensing fields, so this substitution changes the recovered flexion only at higher order. Both estimators can thus be calculated from the PSF-treated observed images within the first-order approximation.


\bibliography{apssamp}{}
\bibliographystyle{aasjournal}



\end{document}